\documentclass[journal=jacsat,manuscript=article,email=true]{achemso}
\usepackage[version=3]{mhchem} 

\usepackage{amssymb}

\usepackage{titlesec}
\titleformat{\section}
    {\normalfont\Large\bfseries}{\thesection.}{1em}{}
\titleformat{\subsection}
    {\normalfont\large\bfseries}{\thesubsection.}{1em}{}

\makeatletter
\renewcommand\thesection{\@arabic\c@section}
\renewcommand\thesubsection{\thesection.\@arabic\c@subsection}
\makeatother
\usepackage[version=3]{mhchem} 

\usepackage{xcolor}
\usepackage[normalem]{ulem} 

\usepackage{multirow}
\usepackage{tablefootnote}
\usepackage{caption}
\usepackage{subcaption}
\usepackage{multicol}
\usepackage{array}
\usepackage{mathtools}

\usepackage{xurl}
\usepackage{hyperref}

\usepackage{graphicx} 
\usepackage{float} 
\usepackage{listings} 
\usepackage{xcolor}
\usepackage{tikz}

\usepackage{upgreek}
\usepackage{pdfpages}

\title{CROSS-MODAL CHARACTERIZATION OF THIN FILM \ce{MoS2} USING GENERATIVE MODELS}
\author{Isaiah A. Moses}
\author{Chen Chen}
\author{Joan M. Redwing}
\affiliation[MRI]{Materials Research Institute, The Pennsylvania State University, University Park, PA 16802}

\author{Wesley F. Reinhart}
\email{reinhart@psu.edu}
\affiliation[MTSE]{Department of Materials Science and Engineering, The Pennsylvania State University, University Park, PA 16802}
\alsoaffiliation[ICDS]{Institute for Computational and Data Sciences, The Pennsylvania State University,
University Park, PA 16802}

\begin{document}

\maketitle

\begin{abstract}
The growth and characterization of materials using empirical optimization typically requires a significant amount of expert time, experience, and resources.
Several complementary characterization methods are routinely performed to determine the quality and properties of a grown sample.
Machine learning (ML) can support the conventional approaches by using historical data to guide and provide speed and efficiency to the growth and characterization of materials.
Specifically, ML can provide quantitative information from characterization data that is typically obtained from a different modality.
In this study, we investigated the feasibility of projecting quantitative metrics from microscopy measurements, such as atomic force microscopy (AFM), using data obtained from spectroscopy measurements, like Raman spectroscopy.
Generative models were also trained to generate the full and specific features of the Raman and photoluminescence spectra from each other and the AFM images of the thin film \ce{MoS2}. 
The results are promising and have provided a foundational guide for the use of ML for the cross-modal characterization of materials for their accelerated, efficient, and cost-effective discovery.
\end{abstract}


\section{Introduction}
The properties and qualities of grown materials are typically determined by extracting information related to different aspects of the materials, including chemical, structural, and morphological properties.
The need for this varied information implies the requirement to use several characterization modalities and instruments.~\cite{li2016synthesis}
Each characterization method gives information generally related to a specific aspect of the materials.
For instance, in Raman spectroscopy, measurement involves the interaction of laser light with the molecular vibrations of materials to provide chemical and structural information.~\cite{mulvaney2000raman, orlando2021comprehensive}
On the other hand, atomic force microscopy (AFM) measurements use a physical probe with a nanoscale tip to measure surface morphology and mechanical properties through force interactions with the sample.~\cite{rugar1990atomic, giessibl2003advances, zhang2018atomic}

These measurements are routinely taken in the epitaxial growth of materials such as thin-film transition-metal dichalcogenides (TMDs).~\cite{hong2021wafer}
TMDs such as \ce{MoS2} are currently receiving significant attention due to their potential applications in various technologies.~\cite{wu2024synthesis, shahbaz2024advancements, ahmed2024recent}. Other applications include biosensors,\cite{lai2025advances} synaptic transistors,\cite{guo2025mos2} and neuromorphic computing.\cite{liu2025mos2}
A standard bottom-up growth method for preparing \ce{MoS2} is metal-organic chemical deposition (MOCVD).
This synthesis method is usually accompanied by several characterization methods, including Raman spectroscopy, photoluminescence (PL), and AFM.~\cite{chen2024effect}
The characteristic bands of Raman spectra include $\mathrm{E^1_{2g}}$, $\mathrm{A_{1g}}$, and $\mathrm{2LA(M)}$ at about 385, 407, and 460 cm$^{-1}$ frequency, respectively.
$\mathrm{E^1_{2g}}$ and $\mathrm{A_{1g}}$ are due to vibrations in-plane and out-of-plane, respectively, while $\mathrm{2LA(M)}$ is associated with longitudinal acoustic phonon.~\cite{li2012bulk, hong2021wafer, liang2018raman, yan2014thermal, lee2010anomalous}.

However, the frequencies and features of these bands vary, for example, with the number of layers and strain in the sample.~\cite{lee2010anomalous}
As the number of layers increases, the two bands move apart due to interlayer vibrations.
In contrast, $\mathrm{E^1_{2g}}$ is blue-shifted (moves toward larger frequency), while $\mathrm{A_{1g}}$ is red-shifted (moves toward smaller frequency) as it becomes a monolayer, with a band split of about 19 cm$^{-1}$ for monolayer reported.~\cite{li2012bulk, catalan2019photoluminescence} 
Other factors affecting the frequency of the bands are the sample temperature and the laser power.~\cite{yan2014thermal, golovynskyi2020exciton}
Similarly, the width and the position of the peak in PL provide information related to the crystal quality, defects, and the number of layers.
For instance, a narrower peak indicates better crystal quality and fewer defects.~\cite{zhang2017shape}
Also, a monolayer normally has a strong photoluminescence emission at room temperature, with the peak around 1.9~eV.
As the film thickness increases from the monolayer to multilayer, the band structure transitions from the direct to the indirect band gap, resulting in the quenching of the intensity.~\cite{chen2024effect}

In contrast to PL, AFM measurement gives morphological features of the sample's surface.
Features that are often determined from the AFM images include surface coverage, grain sizes and density, and roughness of the sample.~\cite{eichfeld2015highly, kandybka2024chemical, moses2024crystal}
Importantly, aside from using different instrumentation and acquiring different information about the samples, these measurements have substantially different latencies.
While Raman and PL measurements can be conducted in the synthesis environment, samples are typically transferred to a separate facility for microscopy measurements.
Furthermore, microscopy measurements generally require a longer duration to complete (as described in the Methods). Additionally, AFM uses tips that become worn out with use, whereas Raman does not require such spare parts.
Notably, while the spectroscopy data used in this study are collected at room temperature after the synthesis is complete, in situ spectroscopy measurements are often performed~\cite{yoo2023review} and have also been reported during MOCVD growth.~\cite{xue2020high}
These differences have important implications for the future of process control and measurement throughput in an increasingly data-driven scientific ecosystem.

In this study, we investigate the deployment of machine learning (ML) to infer information between these measurement modalities.  
Specifically, we evaluate the ability of spectroscopy to provide quantitative data that is typically obtained through microscopy, and vice versa.
The advantages of this approach could substantially accelerate the discovery of new materials, for instance, in determining optimal growth parameters and eliminating the need for various types of equipment for different characterization methods.
Additionally, it would enable a faster estimation of material properties, thereby offering the essential information needed to optimize growth parameters effectively.

Cross-modal learning is a branch of ML that uses information from one modality to estimate data from another modality.
Classical examples of cross-modal learning include deep learning approaches deployed in text-to-speech systems.\cite{kumar2023deep}
In materials and chemical sciences, several studies have deployed cross-modal learning for accelerated materials discovery and characterization.
These include the prediction of molecular structure from the  Nuclear Magnetic Resonance (NMR) spectra,\cite{hu2024accurate} Infrared Spectra,\cite{wu2025transformer, alberts2024leveraging}, and mass spectra.\cite{litsa2023end, stravs2022msnovelist}
These studies have mainly used autoencoder and transformer-based models to elucidate molecular structures from spectra data.
Others have used multi-modal chemical descriptors to train an autoencoder model for the prediction of spectra and structural descriptors,\cite{yang2024cross} the prediction of AFM topography from the optical microscope images using autoencoder models,\cite{jeong2023predicting} and the variational autoencoder model to generate scanning electron microscopy (SEM) images from the corresponding small-angle x-ray scattering (SAXS) 2D pattern and vice versa.\cite{lu2023pair}

In the present study, the availability of both spectroscopic (Raman and PL) and microscopic (AFM) data from \ce{MoS2} samples has paved the way for carrying out both multi- and cross-modal learning.
We performed an unsupervised learning of the AFM to extract a few-dimensional representation from the high-dimensional images.
The few dimensions were observed to be well interpretable, as simple linear correlation shows a strong relationship with physically meaningful properties of the image.
The latent features of the AFM were then learned from the Raman spectra in a successful effort to learn spectroscopy features from microscopy measurements. 
Additionally, generative models were deployed to generate whole spectra of Raman and PL using the corresponding AFM images.
Likewise, Raman spectra were approximated using the PL as input, and vice versa.
A multimodal learning approach, fusing AFM with Raman (PL), was also used to generate PL (Raman).
The generalization of our models established the potential of ML to provide information on samples from one characterization data set that is conventionally obtained from a combination of several characterization modalities.
Importantly, the inference of AFM figures of merit from Raman and PL measurements can be used to rapidly optimize growth conditions, since our AFM instrument has a longer measurement latency compared to our Raman and PL instruments.
Our contribution provides new insights into the capability of ML to accelerate the growth and characterization of TMD materials and beyond.

\section{Methods}

\subsection{Experimental Details}

\subsubsection{\ce{MoS2} Synthesis with MOCVD}
\ce{MoS2} samples included in this study were grown in the 2D Crystal Consortium (2DCC) facility at the Pennsylvania State University.
The growth of \ce{MoS2} monolayer films on 2-inch diameter c-plane sapphire substrates was carried out in a metal-organic chemical vapor deposition (MOCVD) system (instrument DOI: 10.60551/znh3-mj13) equipped with a cold-wall horizontal reactor with an inductively heated graphite susceptor with gas-foil wafer rotation.\cite{zhang2018diffusion}
The MOCVD system features a cold-wall horizontal metalorganic chemical vapor deposition (MOCVD) reactor from CVD Equipment Corp with an inductively heated SiC-coated graphite susceptor that can accommodate a single wafer up to 2 inches in diameter at temperatures up to 1000°C.
Water circulation through the outer quartz jacket maintains a constant wall temperature.
To increase uniformity across the wafer, gas foil rotation of the wafer holder is employed.
The reactor loads and unloads through a nitrogen-purged glovebox to minimize exposure to ambient air.

For \ce{MoS2} monolayer growths, the Molybdenum hexacarbonyl (\ce{Mo(CO)6}) (99.99\%, Sigma-Aldrich) was used as the metal precursor while hydrogen sulfide (\ce{H2S}-99.5\% Praxair) was the chalcogen source.
The \ce{Mo(CO)6} powder was maintained inside a stainless-steel bubbler held at a temperature of 25°C and a pressure of 625~Torr.
Ultra-high-purity hydrogen was used as a carrier gas and was passed through the bubbler at required flow rates to achieve monolayer growth in a comparable duration.
The \ce{MoS2} films were grown in a single-step process described in prior work.\cite{chen2024effect} 

Epi-ready 2-inch c-sapphire (Cryoscore) was used as the substrate for \ce{MoS2} growth.
The substrate was placed on the SiC-coated graphite puck and translated inside the quartz tube, which was subsequently sealed.
The reactor was initially pumped down to a base pressure of $6 \times 10^{-4}$~Torr using a rotary vane pump.
Before the growth, the sapphire substrate was ramped up under \ce{H2} to the growth temperature and pre-annealed for 10~min. The reactor pressure was maintained at 50~Torr throughout the entire \ce{MoS2} growth process.
During growth, \ce{H2S} and \ce{Mo(CO)6} were introduced simultaneously into the inlet gas stream of the reactor for a designated time to complete \ce{MoS2} monolayer growth in a single step.
Then, the \ce{MoS2} monolayer was annealed under \ce{H2} and \ce{H2S} ambient for 10~min at the growth temperature before cooling down to inhibit the decomposition of the obtained \ce{MoS2} film.
The growth process, including growth temperature, \ce{Mo(CO)6} precursor flow rates, and growth time, was carefully designed and optimized for each growth to achieve uniform epitaxial monolayer films with minimal bilayer nucleation.  

\subsubsection{Atomic Force Microscopy Characterization}

Bruker Icon AFM was used to determine the surface morphological features, including the film coverage, domain size, orientations, and thickness at multiple locations on the substrate.

The AFM imaging was performed on a Dimension Icon platform using a Scanasyst-Air probe with a nominal tip radius of approximately 2~nm, a spring constant of 0.4~N/m, and a resonance frequency of about 70~kHz. Height maps of a $5 \mathrm{\upmu m} \times 5 \mathrm{\upmu m}$ and a $1 \mathrm{\upmu m} \times 1 \mathrm{\upmu m}$ area were collected using peak-force tapping mode with a peak-force set point of 1.2~nN. Each image was acquired with a scan resolution of 512 $\times$ 512 pixels and a scan rate of 0.8–1.0Hz, which we selected to ensure optimal tip-sample interaction and feature fidelity across \ce{MoS2} surfaces.

In our study, we used the Height channel data coupled with the Adhesion data from the Dimension Icon AFM system as it provides direct topographic information, which was sufficient for capturing key morphological features such as monolayer and bilayer domains, monolayer coalescence, surface roughness, and step height/terrace width across the \ce{MoS2} films.
Alternative channels, such as the Amplitude Error or “PeakForce Error” signal, can offer enhanced contrast for subtle features, especially in cases with minimal height variation. However, our primary focus was on extracting general surface morphology and domain coverage, which were reliably captured using the Height sensor.

The AFM characterizations were normally carried out in a separate facility upon the completion of the film synthesis process. To acquire AFM images, it typically requires following a dedicated process, including tip swapping, laser alignment, and surface approaching before the scanning. The total scan time per image was 12 to 20 minutes, including automated engagement and image acquisition. As such, acquiring both center and edge scans typically required an hour per 2-inch \ce{MoS2} sample.

AFM data were processed using NanoScope Analysis 2.0 software. To eliminate sample tilt and scanner bow while preserving the integrity of surface features, we applied first-order flattening to remove both constant offset and linear slope generated during the scanning. The height scale bar was adjusted to range from -1.5 nm to +1.5 nm before data export to facilitate direct comparison between different samples.

\subsubsection{Raman and Photoluminescence Spectroscopy}
Room temperature optical mapping using Raman spectroscopy was performed with a Horiba LabRAM HR Evolution spectrometer using 532 nm excitation in a 180$^\circ$ backscattering configuration. A 100X long working distance objective (NA = 0.8) was used, offering a spatial resolution of about $1 \mathrm{\upmu m}$.

To prevent laser-induced heating of the sample, we used a low-power laser with a nominal output of approximately 1.0~mW at the source. Each measurement had an acquisition time of 30 seconds to achieve a suitable signal-to-noise ratio. It is important to note that the actual power measured at the sample's surface, after accounting for losses through the objective and optical path, is about 0.27~mW. To ensure that the sample does not heat from the laser, we monitored for spectral redshift, peak broadening, or any irreversible changes during the acquisition process. None of these effects were observed, indicating a negligible heating effect under our measurement conditions. Additionally, while the sapphire substrate is not as thermally conductive as some metals, it allows for reasonable heat dissipation given the short acquisition times and the lower effective power used.

A spectral grating with 1800 grooves per millimeter was used.
System calibration was performed using single-crystal silicon (520.7~cm$^{-1}$) at room temperature. To account for systematic error, such as room temperature fluctuations, a mercury emission line at 546~nm was monitored. 

PL measurements were performed using the same experimental setup as Raman.
However, a 300 grooves per millimeter spectral grating was used.
The PL spectra were fit to account for spectral contributions from the emission of A-excitons, trions, and defect-bound excitons.
For a single point of Raman and PL spectra acquisition, it normally took a shorter time compared to AFM, about 15 minutes to complete the whole characterization for a 2-inch \ce{MoS2} wafer.

\subsection{Data Preparation}

\begin{figure}
    \centering
    \includegraphics[width=\linewidth]{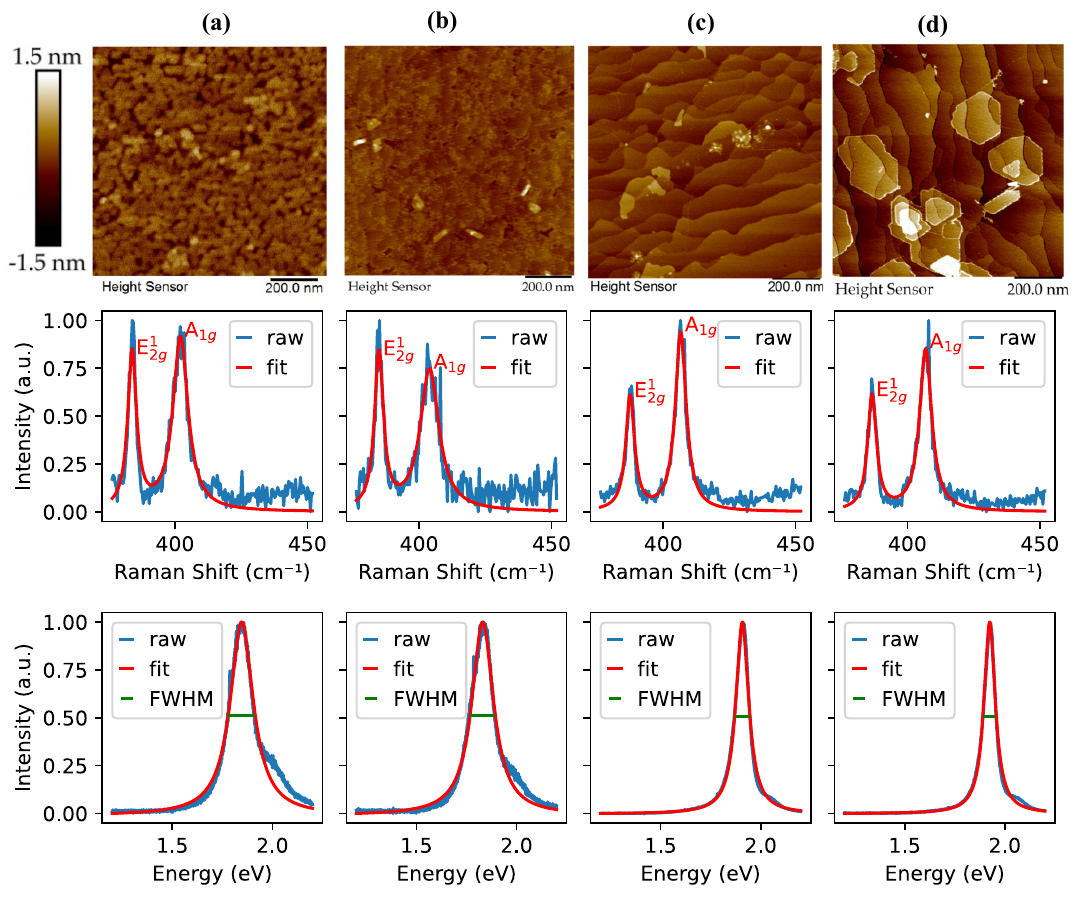}
    \caption{Examples of samples showing the AFM (first row), Raman (second row), and PL (third row). The height scale on the left applies to all the samples.
    Samples \textbf{(a)} and \textbf{(b)} exhibit similarities that are consistent across all characterization modalities (e.g., if they are similar in one characterization technique, they are also similar in others).
    The same pattern holds for samples \textbf{(c)} and \textbf{(d)}.
    To calculate the features of the Raman and PL spectra, such as the $\mathrm{E^1_{2g}}$ and $\mathrm{A_{1g}}$ positions in Raman and FWHM in PL, the data were fitted into the Lorentzian functions.}
    \label{fig:data}
\end{figure}

\begin{table}[]
    \centering
    \begin{tabular}{l| c c   c c} \hline
        Data &   \multicolumn{2}{c}{train and validation}   &   \multicolumn{2}{c}{test}    \\ 
        & unique samples &   total samples    & unique samples & total samples \\ \hline
         AFM-Raman  & 148   &   306   &   27    &   64          \\
         AFM-PL     &   134 &   233  &   25     &   47          \\
         Raman-PL   &   171 &   295   &  24     &   43              \\ 
         AFM+PL - Raman & 131   &   227 &   25  &   44          \\ \hline
    \end{tabular}
    \caption{The data used for training and testing ML models.
    Multiple measurements (such as at the center and edge) for some of the samples result in the total samples column.}
    \label{tab:data}
\end{table}

The data used in this study were retrieved from the LiST hosted by the 2DCC at Pennsylvania State University.\cite{raw_data}
The raw AFM micrographs, Raman spectra, and PL spectra of \ce{MoS2} grown with the MOCVD method with various parameters were recovered. The diverse synthesis parameters of the samples used in the study allow for investigation of the applicability of our models beyond 
a controlled and limited experimental condition. There are multiple measurements for some of the samples (for example, at the center and edge of the wafer).
Hence, the total number of samples is more than the unique number of samples (Table~\ref{tab:data}).
Additionally, for each sample, several sizes of scans were taken from the wafer.
For consistency, only $1 \mathrm{\upmu m} \times 1 \mathrm{\upmu m}$ images were used.

Similarly, the Raman and PL spectra were recorded at multiple laser powers. 
Since the nature of the bands could differ with the laser power,\cite{yan2014thermal} only measurements taken at 4~mW are used. These data cleaning procedures resulted in about 30\% of the original data.
The raw Raman spectra consist of intensities at 1600 different frequencies. However, a physically meaningful part of the Raman spectra are the $\mathrm{E^1_{2g}}$, $\mathrm{A_{1g}}$, and $\mathrm{2LA(M)}$ peaks at about 385, 407, and 460 cm$^{-1}$ frequencies, respectively.~\cite{li2012bulk} 
Therefore, we have used Raman spectra consisting of 202 points in the range containing these important features as input to our ML models.

In holding aside data for testing, we ensured that no samples appeared in both training and validation sets.
In other words, we used grouped splitting with groups defined by the sample identity rather than individual height maps or spectra.
Figure~\ref{fig:data} shows four examples of sample data. The processed data and code used are available for download from Ref\citenum{moses_2025_15533400}.

\subsection{Unsupervised Learning of AFM}

\begin{figure}
    \centering
    \includegraphics[width=\linewidth]{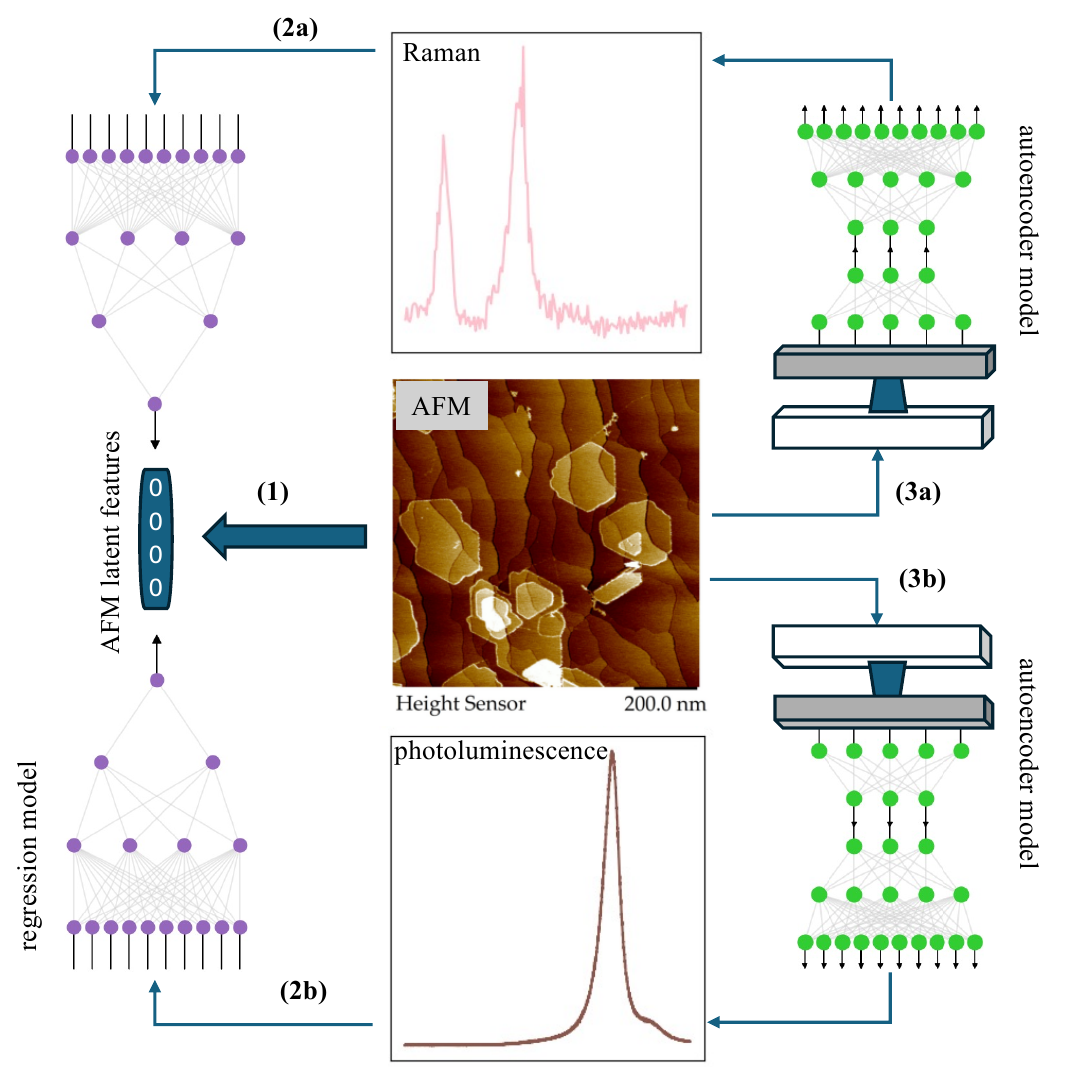}
    \caption{The workflow of the study. \textbf{(1)} is the unsupervised representation of the AFM images to obtain latent features. In obtaining latent features, pretrained ResNet152 is used to extract AFM features and projected to fewer dimensions using Uniform Manifold Approximation and Projection (UMAP) and principal component analysis (PCA). \textbf{(2a)} and \textbf{(2b)} are the regression models of the AFM latent features using the Raman and photoluminescence (PL) spectra, respectively. \textbf{(3a)} and \textbf{(3b)} are the generative models to generate Raman and PL spectra, respectively, from the AFM. Additionally, Raman spectra are generated from PL data (and vice versa) and Raman and PL spectra via AFM fusion with the other spectral type using autoencoder models.}
    \label{fig:models}
\end{figure}

The study started with the reduction of the dimensionality of the AFM images (Figure~\ref{fig:models}).
2048 features were extracted from each image using the average pool layer of ResNet152~\cite{ResNethe2015deep} pre-trained on the ImageNet data.\cite{5206848}
Uniform Manifold Approximation and Projection (UMAP)~\cite{mcinnes2018umap} was then used to reduce the features to 2D.
In UMAP, the number of neighbors, a hyperparameter that determines how the model balances local versus global features in the data, was set to half the sample size (193).
This large number of neighbors ensures that the more global structure in the data is captured, compared to local variation. 
We used the cosine distance and a fixed random state to ensure reproducibility.

There is a wide variety of methods for unsupervised representation learning that may be suitable for this task. In other work, we have found UMAP to be effective in preserving both local and global structure of high-dimensional data.\cite{yu2025multimodal} Without performing a systematic evaluation of all suitable manifold learning methods, we believe UMAP should be as effective as comparable methods like t-distributed Stochastic Neighbor Embedding (t-SNE).\cite{becht2019dimensionality} Ultimately, the exact nature of the learned manifold will be subject to methodology, hyperparameters, and even which data subsets are used.


In addition to UMAP, a linear projection obtained from Principal Component Analysis (PCA) was investigated; the PCA coordinate system yields directions (principal components) representing the highest variance in the data.\cite{PCAdoi:10.1080/14786440109462720, PCAjolliffe2016principal}

To understand the direct physical information captured by UMAP and PCA, we computed a simple pairwise Pearson correlation between the latent and some physically meaningful features of the AFM images.
The physical features examined are the step edges, grain density, root mean square (RMS) roughness, Difference of Gaussian (DoG) blob,\cite{DoGlowe2004distinctive} and dissimilarity, a Gray-Level Co-occurrence Matrix (GLCM)\cite{GLCM4309314} feature.
To determine grains in the images, substrate-crystal segmentation was performed using the Otsu thresholding algorithm~\cite{Otsu4310076} implemented in \texttt{scikit-image} library, where a grain is a connected component of the crystal region.
The algorithm works by maximizing the variance between two classes of pixels to determine the threshold separating the classes (substrate and crystal).
The RMS roughness was determined using the root mean squared error between the smooth Gaussian image and the original image.
To determine the step edges, the Canny edge detection~\cite{Canny4767851} was used, with edges improved using binary dilation and erosion. 
More details on the image analysis implementation parameters can be found in the supporting information of our previous work.\cite{moses2025transfer}

\subsection{Regression Models}

To examine whether microscopy characterization features can be determined using spectroscopic measurements, Raman and PL were used as input features to train regression models to predict the latent features of AFM images.
We used multilayer perceptron (MLP), Ridge, and support vector regression (SVR) models.
MLP was implemented with \texttt{pytorch}\cite{PYTORCHpaszke2019pytorch} while Ridge and SVR were implemented in \texttt{scikit-learn} library version 1.2.2.\cite{SCIKITLEARNpedregosa2011scikit}
For each model, 5-fold cross-validation was used for hyperparameter tuning. The \texttt{optuna} implementation of Bayesian optimization~\cite{optuna_2019} was used to optimize the hyperparameters of MLP.
After each hidden layer of the MLP, we include an activation function and a dropout.
The optimized hyperparameters include the number and size of each hidden layer, the activation function, and the dropout rate applied after each hidden layer.
Others are L1 regularization strength, learning rate, and batch size.

For the Ridge and SVR models, adding up to the second degree and just the first degree, respectively, of the features was observed to yield a better result than simply using the spectra alone.
We did a grid search in the hyperparameter space for the Ridge and support vector regression models. The alpha value was optimized for the Ridge while \texttt{C}, \texttt{degree}, \texttt{gamma}, and \texttt{kernel} were optimized for the SVR.

To further examine the reliability of our regression models, bootstrap confidence interval (CI) analyses were carried out. The coefficient of determination, $R^2$, was used to measure the performance of the models. The CI for $R^2$ was calculated using bootstrap resampling with replacement from the original training dataset. For each bootstrap iteration, models were fitted with the training set using the same hyperparameters obtained from the cross-validation. 30 iterations were used, with the CI calculated using the percentile method from the distribution of $R^2$ across the bootstrap samples.

\subsection{Generative Models}

In addition to regression models, we trained generative models to produce characterization data from a different modality. All generative models utilized autoencoder architectures with fully connected decoder heads. Preliminary investigations indicate that these fully connected decoder heads perform better than more complex architectures, including Long Short-Term Memory (LSTM) networks and transformers (see Table S5).
The encoder architectures were built from scratch, with most parameters optimized using Bayesian optimization implemented in \texttt{optuna}~\cite{optuna_2019} in 50 trials.
For each model, 10-fold cross-validation was used for hyperparameter tuning, with the average validation $\mathrm{R^2}$ score from the 10 models in the cross-validation serving as the metric for evaluating the hyperparameters.
Since the intensity of the Raman and PL was normalized, their values range between 0 and 1.
Therefore, the binary cross-entropy (BCE) loss function was used for the backpropagation in all the generative models trained to generate Raman or PL spectra.

To generate Raman and PL from the AFM, we used pre-trained ResNet152 as the encoder, which our previous study showed to perform well on a similar dataset.\cite{moses2025transfer}
The classification layer of the ResNet152 was replaced with fully connected layers with nodes 2048 and 1024, such that the encoder had a latent dimension of 1024.
The decoder layer was optimized using \texttt{optuna}, including the number of layers, dropout rate, activation function, and learning rate.
We placed batch normalization, activation function, and dropout layer after each inner layer of the decoder, with the number of nodes in each layer et to be half of the preceding layer.
The last layer is the dimension of the Raman and PL for the models trained to generate the respective spectra.
A sigmoid activation function was added to the output layer to ensure the output values are constrained to the range 0 to 1.

Similar to the autoencoders designed to generate Raman and PL from AFM, we also optimized autoencoders to directly predict the $\mathrm{A_{1g}}$-$\mathrm{E^1_{2g}}$ split and FWHM from the Raman and PL, respectively.
The architectures of the autoencoders are the same as those for the entire spectrum, except for the output layers, which output only a scalar without a Sigmoid activation function.
Additionally, MSE was used as the loss function instead of BCE in all other generative models.

In the autoencoder model to generate Raman from PL and vice versa, both the encoder and decoder layers consisted of three fully connected layers with an input and output layer, respectively.
The dimensions of the input and output layers were the same size as the spectra to be encoded and decoded, respectively.
The number of nodes in each layer was optimized within the range 64 to 1024.
The autoencoder architecture is symmetric between the encoder and decoder, with the same number of nodes in the hidden layers, but in reverse order. 
Following each hidden layer is a ReLU activation function, batch normalization, and dropout.
A sigmoid activation function was added to the output layer to ensure the values outputted are only in the range 0 to 1.
Also optimized are the dropout layer, the learning rate, and the batch size.

We also have different autoencoder architectures optimized for multi-modal learning, where a fusion of AFM and Raman or PL spectra is encoded and PL or Raman is decoded.
The encoder consists of the image and spectral encoders.
For the image encoder, which encodes the AFM, a pre-trained ResNet152 model was used, as stated previously.
Fully connected layers were used to encode the spectra (Raman or PL).
After each of the hidden layers of the spectra encoder is a ReLU activation function, batch normalization, and a dropout layer.
We then include a fusion layer, which fuses the features from the image and spectral encoders via concatenation.
Similar to the spectral encoder, the decoder consists of fully connected layers, with each layer followed by a batch normalization, ReLU activation function, and dropout.
The parameters in both the spectra encoder and decoder are optimized.
For the encoder, we optimized the number of layers and the nodes in each layer, while for the decoder, only the number of layers was optimized, and the number of nodes in each layer is half that of the previous layer.
The input size of the decoder is the dimension of the fusion layer, while the output is the dimension of the spectra to be decoded, either Raman or PL.
A sigmoid activation function was added to the output layer to ensure the outputs are within the range of 0 to 1.

\section{Results and Discussion}

\subsection{Unsupervised Learning of AFM}
\begin{table}[ht!]
	\centering
	\caption{The image properties investigated for correlation with the PCA and UMAP embedding obtained from the image features extracted using ResNet pre-trained model. Pearson's correlation coefficient, $R$, is shown here.} 
	\label{tab:property}
	\begin{tabular} {l |c |  c   c    |c  c} \hline
	\multicolumn{2}{c|}{\multirow{2}{*}{\textbf{Property}}} & \multicolumn{4}{c}{\textbf{Correlation} ($R$)} \\ \cline{3-6}
   \multicolumn{2}{c|}{}	 &  UMAP-0  & UMAP-1     & PCA-0 & PCA-1\\ \hline
 1  &  step edges       &  -0.27   &   0.53    &   -0.41    & 0.30 \\
 2  &   grain density   &  0.44    &   -0.61    &   0.59       & -0.24   \\
  3 & rms roughness      & 0.44      &  -0.74    &   0.63    & -0.34  \\
  4  &  DoG blob         & 0.63     &  -0.75       &    0.75    & -0.11    \\ 
 5  &  dissimilarity     &  0.51    &  -0.77    &  0.70       & -0.34     \\ \hline
      
		\end{tabular}
	\end{table}
\begin{figure}
    \centering
    \includegraphics[width=\linewidth]{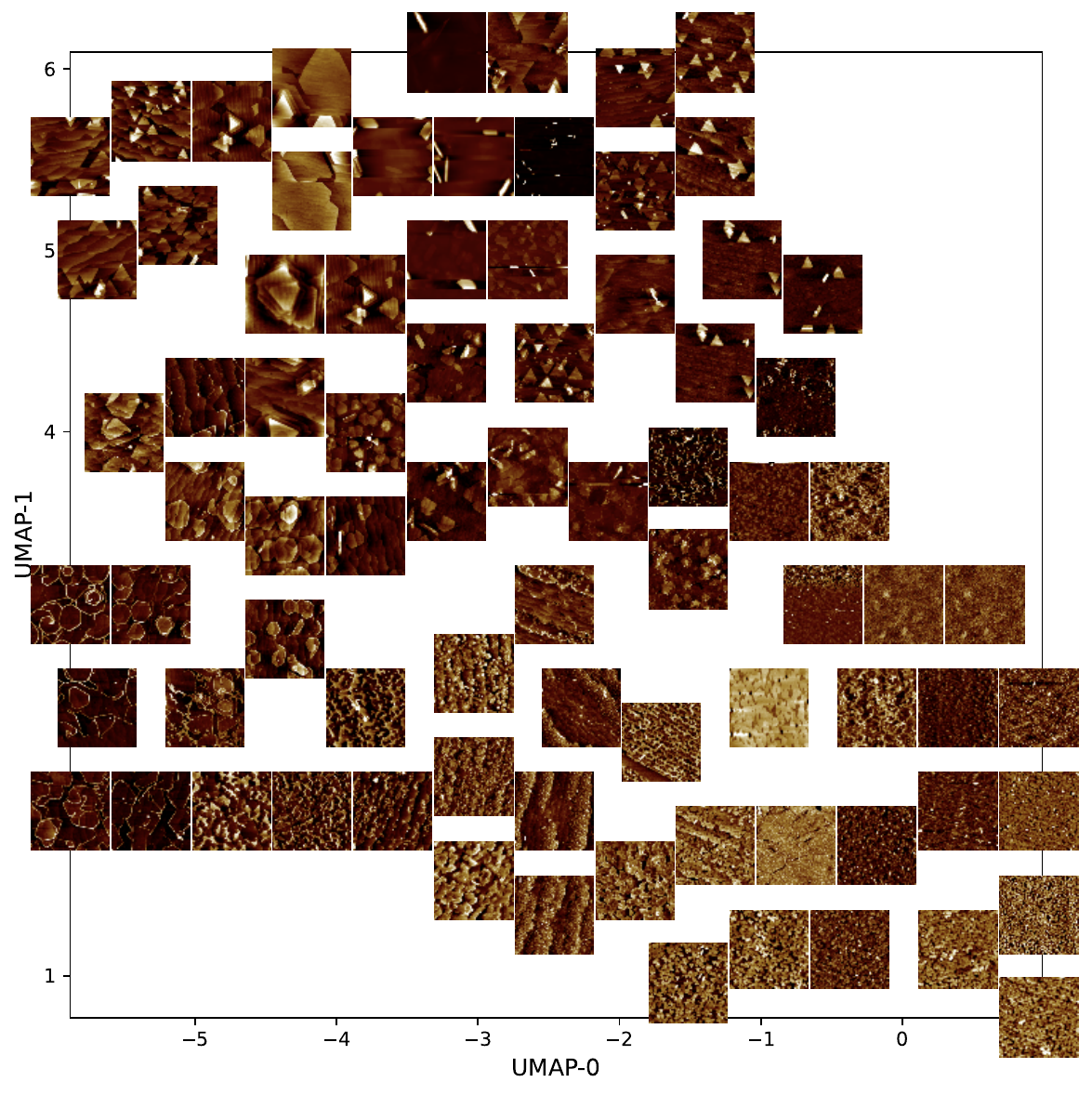}
    \caption{Sample Atomic Force Microscopy (AFM) images in the Uniform Manifold Approximation and Projection (UMAP) embedding space. UMAP-0 and UMAP-1 are the first two components of the UMAP. The figure shows how the features of AFM vary with the components of the UMAP.}
    \label{fig:afm_umap}
\end{figure}

Since there is a relative advantage in the spectroscopy technique, in terms of ease and duration of measurements, compared to microscopy, we wanted to evaluate if AFM features could be learned from Raman spectra.
After obtaining the UMAP and PCA embeddings of AFM features extracted using the ResNet pre-trained model, we interpreted the latent features in relation to physical image features.
Sample AFM images were therefore embedded in the UMAP (Figure~\ref{fig:afm_umap}) and PCA (Figure~S1) space.
All the AFM images are available online with our data package.\cite{moses_2025_15533400}

The UMAP embedding reveals a continuous and physically meaningful organization of \ce{MoS2} surface morphologies across the embedding space (Figure~\ref{fig:afm_umap}).
Furthermore, the two UMAP components map well to what is known about \ce{MoS2} crystal growth. 
The horizontal axis, UMAP-0, captures the characteristic feature size, progressing from large, well-defined crystalline domains on the left (negative UMAP-0) to fine-grained, polycrystalline films on the right (positive UMAP-0). 
This is quantitatively supported by the positive correlation of UMAP-0 with DoG blob ($R=0.63$) and grain density ($R=0.44$) in Table~\ref{tab:property}, metrics that increase with the number of small features.
The vertical axis, UMAP-1, distinguishes the degree of structural order and surface quality, ranging from disordered morphologies at the bottom (negative UMAP-1) to smooth, highly ordered surfaces at the top (positive UMAP-1).
This axis is anti-correlated with metrics of disorder, including RMS roughness ($R=-0.74$) and dissimilarity ($R=-0.77$), while showing a positive correlation with step edges ($R=0.53$), confirming that positive UMAP-1 values represent the smooth, terraced surfaces characteristic of high-quality epitaxial growth~\cite{dumcenco2015large}.

By considering these two axes together, the embedding organizes the AFM images into a range of physically relevant archetypes that reflect known \ce{MoS2} growth mechanisms. 
The top-left corner is populated by large, faceted triangular and hexagonal domains with flat terraces.
These are characteristics of high-quality, single-crystal monolayer \ce{MoS2} grown under optimal CVD conditions with low nucleation density~\cite{dumcenco2015large, venkata2016atomically}.
Conversely, the bottom-right corner contains fine-grained, rough, and disordered surfaces, indicative of polycrystalline films formed under conditions of high nucleation density. 
Such films are a known outcome of CVD processes where growth parameters induce defects~\cite{venkata2016atomically}.
The bottom-left region shows large but more irregular or coalesced domains, suggesting defective or incomplete monolayer growth.
The prevalence of triangular versus hexagonal flakes, particularly visible on the left side of the embedding, reflects the relative stability of Mo- and S-terminated edges, which is highly sensitive to the sulfur and hydrogen chemical potentials during synthesis~\cite{lauritsen2004atomic}.

Thus, the UMAP embedding not only organizes the images by visual similarity but also maps them onto a landscape defined by known modes of \ce{MoS2} growth, from ideal van der Waals epitaxy to defective polycrystalline films.
To quantify this interpretation, Pearson's correlation coefficients were calculated between the latent features and the physical image features (Table~\ref{tab:property}).
The strong correlations between the UMAP components and properties like RMS roughness, grain density, and feature dissimilarity validate the physical meaning of the embedding axes.
Given that the UMAP components show stronger and more distinct correlations with these physical features than the PCA components, we select only UMAP for all subsequent analysis.

\subsection{Learning Microscopy Features from Spectroscopy Modality}

\begin{table}[ht!]
    \centering
    \begin{tabular}{l   |   c   c   c  | c   c   c | c  c   c}\\ \hline
        \textbf{Embedding} &\multicolumn{3}{c|}{\textbf{train} $R^2$} &   \multicolumn{3}{c|}{\textbf{test} $R^2$}&  \multicolumn{3}{c}{\textbf{test} 95\% CI}\\ 
                &   MLP     &   Ridge   &   SVR     &   MLP     &   Ridge   &   SVR &   MLP     &   Ridge   &   SVR    \\ \hline
        UMAP-0  &   0.76    &   0.73    &   0.75    &   0.34    &   0.59    &   0.56&   [-0.05, 0.43]&  [0.92, 1.00] & [0.66, 0.83]    \\
        UMAP-1  &   0.81    &   0.72    &   0.81    &   0.54    &   0.68    &   0.69&   [0.11, 0.67]  &   [0.93, 1.00]  &   [0.69, 0.84]    \\ \hline

    \end{tabular}
    \caption{Model performance in predicting the first two components of UMAP embedding (UMAP-0 and UMAP-1) from the Raman spectra. The first two columns show the coefficient of determination ($R^2$) on the train and test sets, respectively. The last column shows the 95\% confidence interval (CI) for the $R^2$}
    \label{tab:raman_afm}
\end{table}

\begin{figure}
    \centering
    \includegraphics{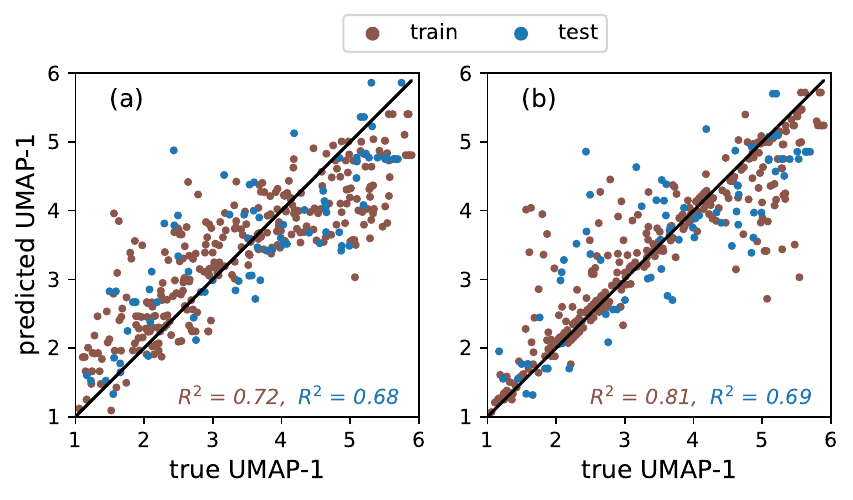}
    \caption{Predicting the AFM embedding from the Raman spectra. (a) and (b) are the second components of UMAP from the Ridge and SVR, respectively.}
    \label{fig:raman_afm}
\end{figure}
\clearpage

Having established the validity of using latent features to measure the physically meaningful properties of the AFM, we then further explored the latent features of the AFM using Raman and PL spectra.
Regression models were trained to predict the latent characteristics of AFM.
The optimal hyperparameters obtained from Bayesian optimization of the MLP and the grid search of the Ridge and SVR models are available in the Supporting Information.
The results show that Raman spectra can be used to predict these characteristics with reasonable accuracy (Table~\ref{tab:raman_afm} and Figure~\ref{fig:raman_afm}).
Specifically, MLP, Ridge, and SVR all show that we can better learn UMAP-1  from the Raman spectra (Table \ref{tab:raman_afm}).
The MLP exhibits the least generalization capacity among the three regression models, with coefficient of determination ($R^2$) values of 0.34 and 0.54 in the test set, and 0.76 and 0.81 in the training set for UMAP-0 and UMAP-1, respectively.
Ridge provides significantly better generalization, with $R^2$ values of 0.59 and 0.68 for UMAP-0 and UMAP-1, respectively.
In addition, while SVR is better on the UMAP-1 with $R^2$ value of 0.69, its performance is lower than Ridge on the UMAP-0 with an $R^2$ value of 0.56. 
In predicting the UMAP-0, the CI widths are 0.48, 0.08, and 0.17 for MLP, Ridge, and SVR, respectively (Table \ref{tab:raman_afm}). Similarly, in predicting the UMAP-1, the CI widths are 0.56, 0.07, and 0.15. The Ridge model has therefore shown a very high precision for the $R^2$ estimation while SVR and MLP give moderate and low precisions, respectively.
The scatter plots of true and predicted UMAP-1 from the best-performing models, Ridge and SVR, are shown in Figure~\ref{fig:raman_afm}). 

Interestingly, the regression models used to predict the AFM physical features based on the Raman spectra produced poor results (see Table S3). While the $R^2$ values for the training set were high for some features, their generalization to the test sets was weak. This difficulty in predicting physical features from the spectra may stem from the Raman signals not accurately representing any of the quantified physical features from the images. Instead, the latent features appear to capture broader properties of the images, which is why their prediction using the Raman spectra yielded better results.

In contrast to the Raman spectra, PL failed to yield a viable predictive model for AFM latent features (Table~S4).
This might likely stem from the inherent information density differential between the spectroscopic techniques.
While PL is a convolution of mostly only two exciton peaks (A exciton and trion), with possible additional peaks such as the B and defect bond excitons at higher and lower energies, respectively, the Raman spectra contain multiple characteristic vibrational modes that encode structural and crystallographic information, including defect density, strain, and layer interaction.\cite{liang2018raman}
Most fundamentally, the Raman provides a direct measurement of vibrational modes in the material structure, which are more sensitive to the morphological variations detectable by AFM.\cite{wu2023analyzing}
In contrast, PL signals represent electronic transitions that may be less directly influenced by surface features and more affected by bulk properties, defect states, and adsorbates on the surface.\cite{xu2019analysis}
Common information contained in Raman spectra and AFM, which we have confirmed from our ML analysis, includes surface roughness, grain density, DoG blob size, and image morphology dissimilarity (Table~\ref{tab:property}).

The analysis indicates that the UMAP dimension corresponding to the most critical aspects of crystal quality is indeed the one most accurately predicted from the Raman spectra.
As established in our analysis of Figure~\ref{fig:afm_umap}, UMAP-1 distinguishes the degree of structural order and surface quality, ranging from disordered to highly ordered surfaces (validated via anti-correlations with RMS roughness and dissimilarity).
This dimension is also the most predictable, achieving a test $R^2$ of up to 0.69.
In contrast, UMAP-0, which primarily captures feature size, is predicted with lower accuracy (test $R^2$ up to 0.59).

This result demonstrates that the regression model has learned a genuine physical relationship, as the Raman spectra are most sensitive to the exact morphological characteristics encoded by UMAP-1.
Furthermore, the continuous organization of morphologies in Figure~\ref{fig:afm_umap} shows that this mapping is robust; samples with significantly different embedding coordinates are unlikely to share the same physically derived geometric characteristics.
Altogether, this analysis establishes a quantitative link between the AFM and Raman modalities, validating the use of spectroscopy for the rapid estimation of morphological features.

The approach discussed could be applied to other materials, and leveraging transfer learning from \ce{MoS2} to different materials may offer a significant advantage. Additionally, we anticipate improved generalizations with larger datasets. For example, we have previously compared the performance of deep learning across different materials classes and found that incorporating a larger volume of data enhances performance.\cite{moses2025transfer} While the advantages of transfer learning have been demonstrated in AFM for both single and multi-material instances,\cite{moses2024crystal, moses2024quantitative} it remains to be seen what benefits can arise from using Raman and PL spectroscopy data in a transfer learning context.

\subsection{Generative Cross-Modal Learning}

The previous sections have established connections between properties measured using different characterization modalities, Raman and AFM.
Beyond predicting a specific feature, generative models can estimate the entire spectrum from other modalities.
Here, we generate Raman signals from AFM and PL, and merge AFM and PL using autoencoder models.
Similarly, PL spectra are generated from AFM, Raman, and the combination of AFM and Raman.
The hyperparameters obtained from the Bayesian optimization for each model are given in the Supporting Information.

\begin{figure}
    \centering
    \includegraphics{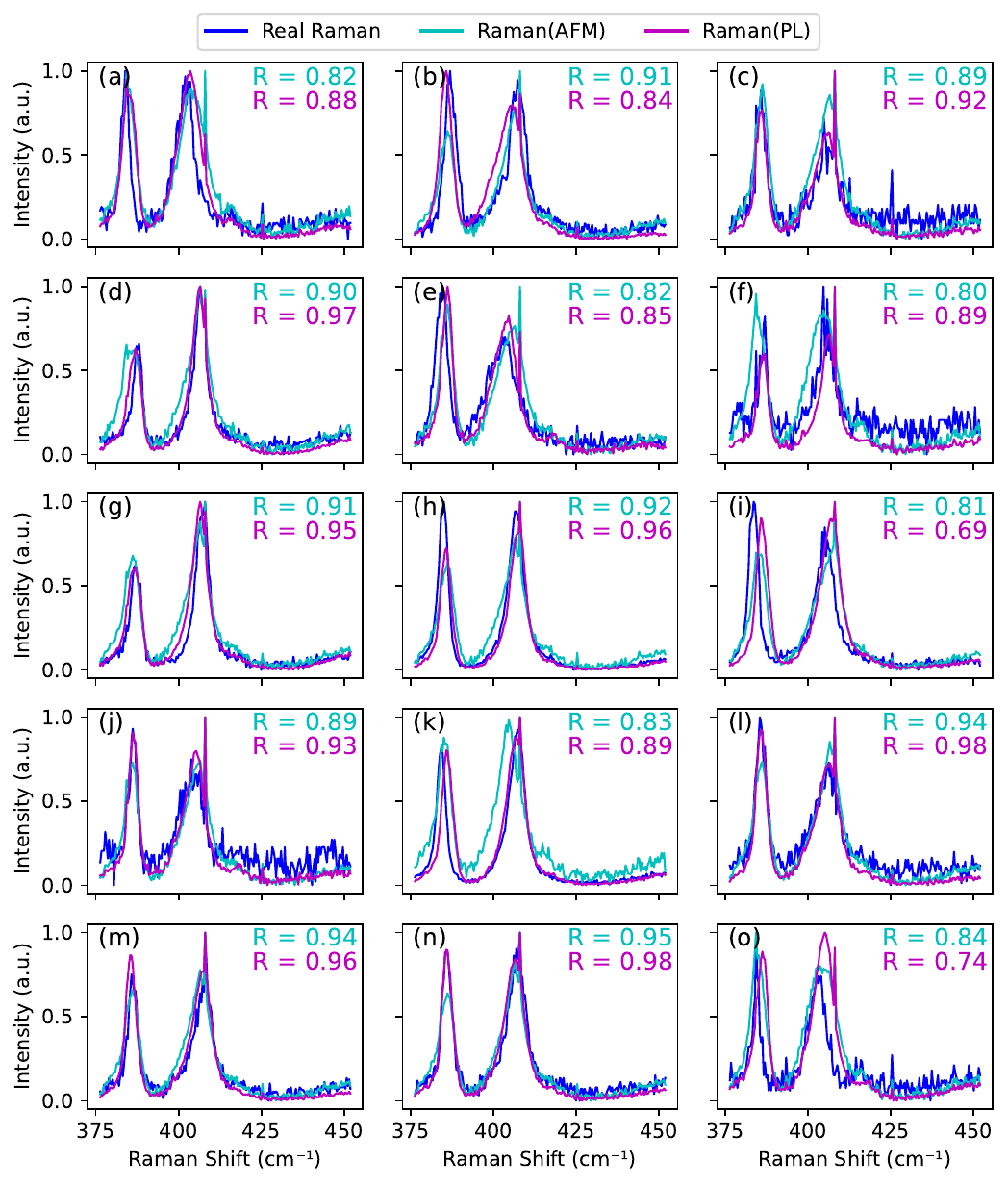}
    \caption{Sample held-out Raman spectral generated from the corresponding AFM images and PL spectra. The Pearson's correlation coefficient, $R$, is between the real Raman spectrum and the Raman spectrum generated from the AFM and PL, respectively, as indicated by the color.}
    \label{fig:raman_generated}
\end{figure}

\begin{table}[ht]
    \centering
    \begin{tabular}{l|c   c |  c   c |  c   c   }  \hline
         &  \multicolumn{2}{c|}{\bf{Raman(AFM)}} & \multicolumn{2}{c|}{\bf{Raman(PL)}} &\multicolumn{2}{c}{\bf{Raman(AFM,PL)}}    \\
         & RMSE    &   $R$   &   RMSE    &   $R$   & RMSE  &   $R$     \\ \hline
        train   & 0.06$\pm$0.01  & 0.97$\pm$0.00 & 0.10$\pm$0.00  & 0.92$\pm$0.01  &   0.09$\pm$0.01  & 0.93$\pm$0.01     \\     
        val     & 0.07$\pm$0.00 & 0.96$\pm$0.01  & 0.11$\pm$0.01 &  0.90$\pm$0.02  &   0.11$\pm$0.01   &   0.90$\pm$0.02   \\
        test    & 0.12$\pm$0.00 & 0.89$\pm$0.00  &   0.12$\pm$0.00    &   0.88$\pm$0.01  &   0.12$\pm$0.01   &   0.89$\pm$0.01        \\ \hline
    \end{tabular}
    \caption{The performance of models on cross-generation of Raman spectra. Raman(AFM) is generated from AFM only, Raman(PL) is generated from PL only, and Raman(AFM,PL) is generated from the fusion of AFM and PL.}
    \label{tab:raman_generated}
\end{table}

Samples of generated Raman spectra from the held-out test sets are shown in Figure~\ref{fig:raman_generated}.
This yields excellent results, with the model capable of generating the most important peaks, the $\mathrm{A_{1g}}$ and $\mathrm{E^1_{2g}}$, as well as realistic noise.
Additionally, the artifacts from the samples, as shown in the measurements, are mostly reproduced by the models, for example, in panels \ref{fig:raman_generated} (a), (b), and (e). However, subtle differences in peak width, baseline offset, and noise structure are not completely captured.
On the train and validation sets, the accuracy of the Raman generated from the AFM input modality is better than that of the PL and merged AFM and PL inputs (Table~\ref{tab:raman_generated}).
However, the RMSE and $R$ on the test set are the same across the three models, indicating that this may not be true in general.

Using AFM input only provides additional benefits in real applications, as both Raman and PL measurements are typically taken together (i.e., from the same instrument).
In contrast, AFM measurements are performed on a separate instrument, which better motivates the need for cross-modal learning.
Spectra from the merged AFM and PL inputs are shown in the Supporting Information; the accuracy is quite similar to that obtained from using PL alone.
The performance in generating Raman spectra from the merged AFM and PL inputs is not distinguishable compared to AFM alone, possibly as a result of the more complex model and smaller data set (Table~\ref{tab:data}) available for the former.

\begin{figure}
    \centering
    \includegraphics{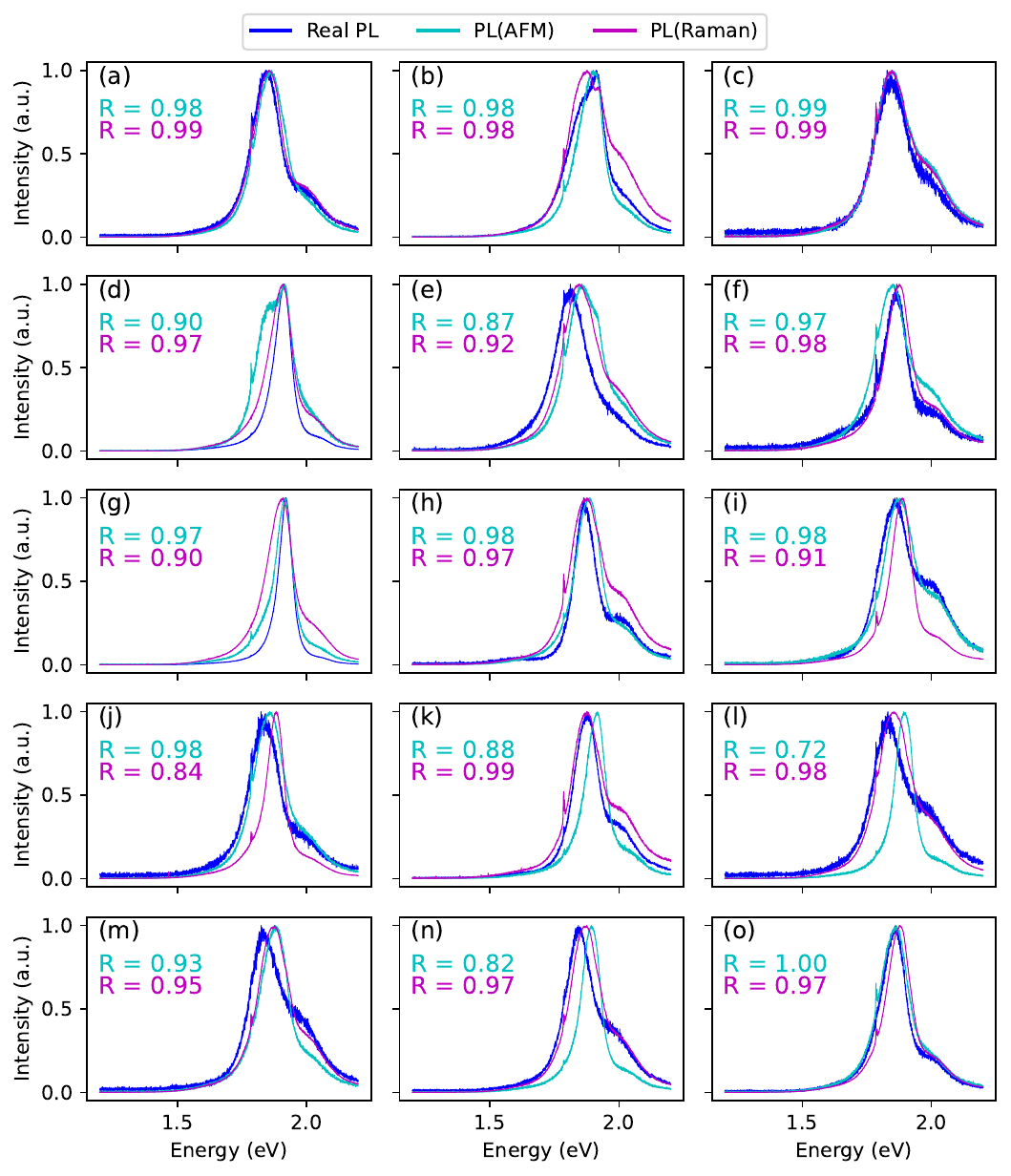}
    \caption{Sample held-out photoluminescence (PL) spectra generated from the corresponding AFM images and Raman spectra. The Pearson's correlation coefficient, $R$, is between the real PL spectrum and the PL spectrum generated from the AFM and Raman, respectively, as indicated by the color.}
    \label{fig:pl_generated}
\end{figure}

\begin{table}[ht]
    \centering
    \begin{tabular}{l|c   c |  c   c |  c   c   }  \hline
         &  \multicolumn{2}{c|}{\bf{PL(AFM)}} & \multicolumn{2}{c|}{\bf{PL(Raman)}} &\multicolumn{2}{c}{\bf{PL(AFM,Raman)}}    \\
         & RMSE    &   $R$   &   RMSE    &   $R$   & RMSE  &   $R$     \\ \hline
        train   & 0.04$\pm$0.00  & 0.99$\pm$0.00  & 0.05$\pm$0.01& 0.98$\pm$0.01  &    0.05$\pm$0.01 &  0.99$\pm$0.01    \\     
        val     & 0.06$\pm$0.02   &   0.98$\pm$0.02  &  0.08$\pm$0.01   & 0.95$\pm$0.01  & 0.08$\pm$0.01    & 0.96$\pm$0.01     \\
        test    & 0.08$\pm$0.01   &   0.95$\pm$0.01   &  0.09$\pm$0.01   & 0.95$\pm$0.01   &  0.08$\pm$0.00    &  0.96$\pm$0.01        \\ \hline
    \end{tabular}
    \caption{The performance of models on cross-generation of PL spectra. PL(AFM) is generated from AFM only, PL(Raman) is generated from Raman only, and PL(AFM,Raman) is generated from the fusion of AFM and Raman.}
    \label{tab:pl_generated}
\end{table}

Similar to the Raman spectra generation, samples of PL spectra generated from the test set are shown in Figure~\ref{fig:pl_generated}, and the correlation between the real and generated PL is shown in Table~\ref{tab:pl_generated}.
The models mostly produce accurate peak positions and intensities.
Additionally, the widths and the peculiar inflection regions of the spectra are accurately produced in some of the samples using both the AFM and Raman (Figure~\ref{fig:pl_generated} (a), (c), and (o)).
Some are better when generated using the AFM (Figure~\ref{fig:pl_generated} (f), (h), and (i)) while the Raman seems to perform better in others (Figure~\ref{fig:pl_generated} (k) and (i)).
There doesn't seem to be a clear difference as to the nature of spectra that the AFM or the PL better generates.
Similar results are obtained with fused AFM and Raman inputs (Figure~S3).

\begin{table}[ht]
    \centering
    \begin{tabular}{l|c     c   c | c   c   c  }  \hline
    &   \multicolumn{3}{c|}{\textbf{$\mathrm{A_{1g}}$-$\mathrm{E^1_{2g}}$ split}}    &   \multicolumn{3}{c}{\textbf{FWHM}} \\
         &  Raman(AFM) & Raman(PL) & Raman(AFM,PL) & PL(AFM)   & PL(Raman)   & PL(AFM,Raman)    \\ \hline
       
        train     & 0.88 &  0.65  &   0.71 &   0.94 &  0.89 &   0.84  \\     
        val      & 0.97  &   0.73  &   0.33 &   0.97    & 0.56   &   0.71   \\
        test     & 0.52  &    0.58  &   0.31 &   0.23   &  0.29  &  0.32   \\ \hline
    \end{tabular}
    \caption{The $R^2$ value of models on predicting the $\mathrm{A_{1g}}$-$\mathrm{E^1_{2g}}$ split and FWHM from the generated Raman and PL spectra, respectively.
    Raman(AFM) is generated from AFM only, Raman(PL) is generated from PL only, and Raman(AFM,PL) is generated from the fusion of AFM and PL.
    PL(AFM) is generated from AFM only, PL(Raman) is generated from Raman only, and PL(AFM,Raman) is generated from the fusion of AFM and Raman.}
    \label{tab:spectra_features}
\end{table}

\begin{figure}
    \centering
    \includegraphics[width=0.75\linewidth]{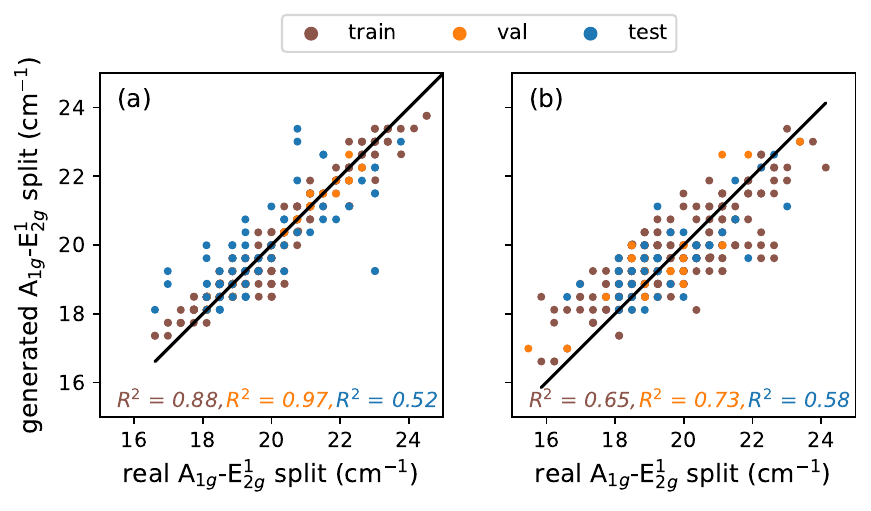}
    \caption{$\mathrm{A_{1g}}$-$\mathrm{E^1_{2g}}$ split obtained from the Raman generated from (a) AFM images and (b) PL spectra.}
    \label{fig:raman_split}
\end{figure}

Beyond generating full spectra, we aimed to determine the quantitative features of the generated spectra and compare them with those obtained from true spectra.
The important features, $\mathrm{A_{1g}}$-$\mathrm{E^1_{2g}}$ split from Raman and FWHM from PL, were obtained from both the true and the generated spectra.
Table~\ref{tab:spectra_features} reports the $R^2$ between the features from the true and generated spectra.
The features obtained from spectra generated by AFM inputs show superior performance on validation data, but generalization to the test set is better with Raman from PL, and PL from fused AFM and Raman.
The estimation of Raman features is shown to be superior from AFM images and the PL spectra, resulting in predictive models with $R^2$ values of 0.52 and 0.58, respectively (Table~\ref{tab:spectra_features}, Figure~\ref{fig:raman_generated}).

\section{Conclusion}

This study presents a robust machine learning framework for cross-modal material characterization, specifically applied to thin film \ce{MoS2}.
By integrating unsupervised representation learning, supervised regression models, and self-supervised generative models, we successfully demonstrated the ability to infer information about the materials across three characterization modalities.
Our unsupervised embedding of AFM images yielded latent dimensions with strong correlations to physically meaningful features such as surface roughness, grain density, DoG blob size, and image dissimilarity.
We then demonstrated that these microscopy-derived latent features can be predicted accurately from corresponding Raman spectra using supervised regression models (MLP, Ridge, and SVR), establishing a quantifiable link between spectroscopy and microscopy.

Furthermore, our trained generative models successfully reconstructed full Raman and PL spectra from AFM image inputs and demonstrated the ability to convert between Raman and PL spectra.
The strong agreement between quantitative features, $\mathrm{A_{1g}}$-$\mathrm{E^1_{2g}}$ split from the Raman and FWHM from the PL, extracted from the generated spectra and the corresponding measured spectra, validates the scheme's flexibility.
The accuracy of the models in generating the full spectra and their important quantitative features, despite the limited data, is promising. 
This establishes the potential of ML in minimizing the cost and complications from employing multiple instruments for materials growth and characterization during synthesis parameter optimization.
The implementation of such generative models for real-world applications will provide efficiency, save time, and accelerate the optimization of material growth parameters.

However, despite the potential of cross-modal learning, our work also reveals the challenges in training generative models on limited datasets.
The LiST database, populated by dozens of researchers within the 2DCC at Penn State, already represents a large-scale database in the context of conventional materials science.
Acquiring much more data would likely require new approaches such as high-throughput methods, including closed-loop autonomous systems.
Ultimately, we believe the methods developed here represent foundational work that can be expanded in a new era of materials science driven by artificial intelligence and automation.

\section*{Acknowledgments}
This study is based on research conducted at the Pennsylvania State University Two-Dimensional Crystal Consortium—Materials Innovation Platform (2DCC-MIP), which is supported by the NSF cooperative agreement DMR-2039351.

\section*{Data Availability}
The raw data required to reproduce these findings can be downloaded from Ref\citenum{raw_data}.
The processed data and code required to reproduce these findings can be downloaded from Ref\citenum{moses_2025_15533400}.

\clearpage
\bibliography{main}


\begin{figure}
\textbf{Table of Contents}\\
\medskip
  \includegraphics{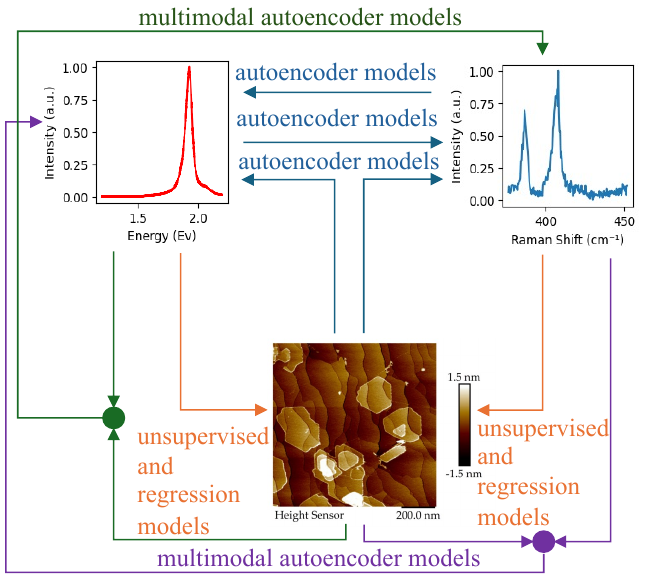}
  \medskip
  \caption*{Cross-modal learning is evaluated using atomic force microscopy (AFM), Raman spectroscopy, and photoluminescence spectroscopy (PL) through unsupervised learning, regression, and autoencoder models. Autoencoder models are used to generate spectroscopy data from the microscopy images. The analysis also involves data fusion techniques, combining AFM and Raman data to generate PL, and merging AFM with PL data to produce Raman spectra.}
\end{figure}

\clearpage


\includepdf[pages=-]{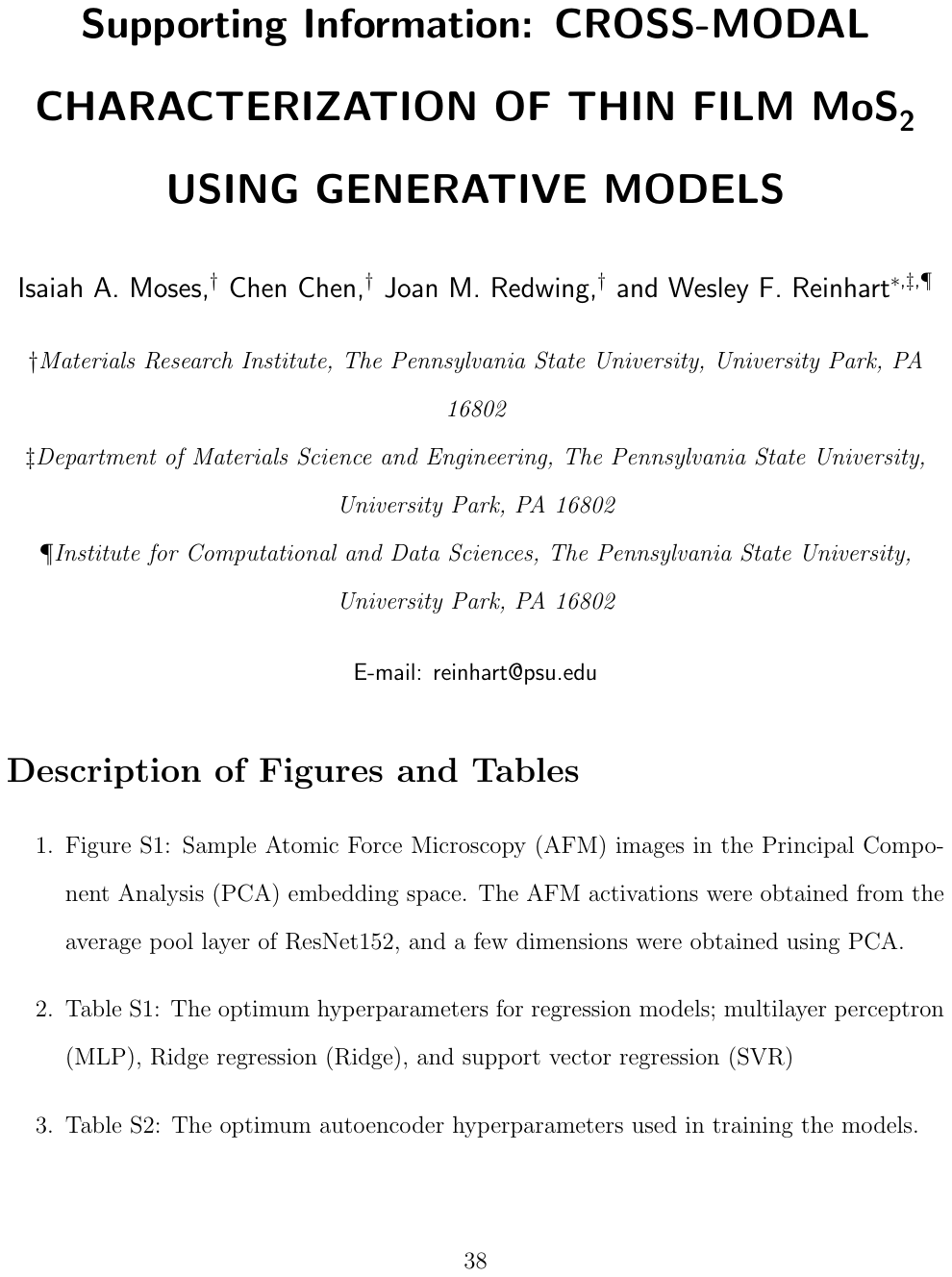}
\end{document}